\documentclass{ws-ijmpd}
\usepackage[super,compress]{cite}
\begin{document}

\markboth{Alicia L\'opez-Oramas, Oscar Blanch, Emma~de O\~na Wilhelmi et al.}
{VHE observations of binary systems performed with the MAGIC telescopes}

%
\catchline{}{}{}{}{}
%

\title{VHE OBSERVATIONS OF BINARY SYSTEMS PERFORMED WITH THE MAGIC TELESCOPES}

\author{Alicia L\'opez-Oramas$^{1, 2}$}

\address{\it $^1$\small Inst. de Astrof\'isica de Canarias, E-38200 La Laguna, Tenerife, Spain\\
\it $^2$\small Universidad de La Laguna, Dpto. Astrof\'isica, E-38206 La Laguna, Tenerife, Spain\\
aloramas@iac.es}

\author{Oscar Blanch$^{3}$, Emma~de O\~na Wilhelmi$^{4}$, Alba~Fern\'andez-Barral$^{3}$, Daniela~Hadasch$^5$, Elena~Moretti$^6$, Pere Munar-Adrover$^7$, Josep~Maria~Paredes$^8$, Marc~Rib\'o$^8$, Diego~F.~Torres$^4$ (the MAGIC Collaboration), Pol Bordas$^9$, Fran\c{c}ois Brun$^{10}$, Jorge Casares$^{1,2}$ and Roberta Zanin$^9$}

\address{{\it $^3$\small Institut de Fisica d'Altes Energies (IFAE), The Barcelona Institute\\
\small of Science and Technology, Campus UAB, 08193 Bellaterra (Barcelona), Spain}\\
{\it $^4$\small Institute for Space Sciences (CSIC/IEEC), E-08193 Barcelona, Spain}\\
{\it $^5$\small Japanese MAGIC Consortium, ICRR, The University of Tokyo,\\
\small Department of Physics and Hakubi Center, Kyoto University,\\
\small Tokai University, The University of Tokushima, Japan}\\
{\it $^6$\small Max-Planck-Institut f\"ur Physik, D-80805 M\"unchen, Germany}\\
{\it $^7$\small Unitat de F\'isica de les Radiacions, Departament de F\'isica,\\
\small and CERES-IEEC, Universitat Aut\`onoma de Barcelona, E-08193 Bellaterra, Spain
}\\
{\it $^8$\small Universitat de Barcelona, ICC, IEEC-UB, E-08028 Barcelona, Spain}\\
{\it $^9$\small Max-Planck-Institut f\"ur Kernphysik, P.O. Box 103980, D 69029 Heidelberg, Germany }\\
{\it $^{10}$\small Universit\'e Bordeaux, CNRS/IN2P3, Centre d'\'Etudes\\
\small Nucl\'eaires de Bordeaux Gradignan, 33175 Gradignan, France}\\}

\maketitle

\begin{history}
\received{16 February 2018}
\accepted{20 February 2018}
\end{history}

\begin{abstract}
The improvement on the Imaging Air Cherenkov Technique (IACT) led to the discovery of a new type of sources that can emit at very high energies: the gamma-ray binaries. Only six systems are part of this exclusive class. We summarize the latest results from the observations performed with the MAGIC telescopes on different systems as the gamma-ray binary LS I +61$^{\circ}$ 303 and the microquasars SS433, V404 Cygni and Cygnus X-1, which are considered potential VHE gamma-ray emitters. The binary system LS I +61$^{\circ}$ 303 has been observed by MAGIC in a long-term monitoring campaign. We show the newest results of our search for super-orbital variability also in context of contemporaneous optical observations. Besides, we will present the results of the only super-critical accretor known in our galaxy: SS 433. We will introduce the VHE results achieved with MAGIC after 100 hours of observations on the microquasar Cygnus X-1 and report on the microquasar V404 Cyg, which has been observed with MAGIC after it went through a series of exceptional X-ray outbursts in June 2015.\end{abstract}




\section{Introduction}	

The development of the new generation of IACTs and improved analysis techniques led to the discovery of a new class of binaries: the gamma-ray binaries. These systems emit high energy (HE:0.1-100 GeV) and/or very high energy (VHE$\ge$100 GeV) gamma rays and their non-thermal emission peaks beyond 1 MeV in a spectral luminosity diagram. They are composed by a massive star and a compact object, either a neutron star (NS) or a black hole (BH). The discovery of this new type of binaries provide a new window to the study of particle acceleration, accretion (and ejection) processes and magnetized relativistic outflows. Up to now, only six systems can be classified as gamma-ray binaries, and the uncertainties on the nature of the compact object and the scenario to explain the gamma-ray emission still remain. 

In order to address these questions, MAGIC has searched for VHE emission from binary systems for over a decade. The MAGIC telescopes are two IACTs of 17\,m-diameter located on La Palma (Canary Islands, Spain), at the observatory of El Roque de Los Muchachos (28$^\circ$N, 18$^\circ$W, 2200 m above the sea level). The integral sensitivity above 290 GeV is (0.67 $\pm$ 0.04 )\% of the Crab Nebula flux in 50 hours \cite{Aleksic2016}. In this contribution, we summarize the latest results from the observations performed with the MAGIC telescopes on different compact binaries.


\section{The gamma-ray binary LS~I~+61$^{\circ}$303}

LS~I~+61$^{\circ}$303 is one of the few binary systems detected from radio to {VHE gamma rays}. It is composed of a B0Ve \cite{Hutchings81} star  with a circumstellar disk and a compact object of unknown nature. It was first detected at VHE by MAGIC in 2006 \cite{Albert_Science_LSI}. It shows an orbital period of 26.496 days and a super-orbital period of 1667 $\pm$ 8 days detected in radio \cite{Paredes1987, Gregory02} and confirmed in optical and HE gamma-rays \cite{Fermi2013}. At Fermi-LAT energies, the super-orbital variability is almost invisible around the periastron, where the compact object is inside (or highly affected by) the Be circumstellar disk, but appears around apastron. This behavior can be explained in a transitioning pulsar-wind scenario \cite{Zamanov2001} like the flip-flop model\cite{Torres12} where changes in the Be star mass-loss rate can cause the switching from propeller regime (at periastron) to an ejector regime (at apastron), when particles are accelerated to TeV energies in the inter-wind shock.

Similarly, we have searched for super-orbital modulation at VHE. For that purpose, we have performed observations between 2010 and 2014 (during the apastron phases, $\phi$=0.5-0.75) and we have included archival (published) MAGIC and VERITAS data, to increase the statistics.  We have detected super-orbital variability in the TeV peak of LS~I~+61$^{\circ}$303 \cite{Ahnen2016} compatible with the 1667-day radio modulation at 8\% probability, assuming a sinusoidal function. The long-term modulation is shown in Fig.~\ref{f1} . The temporal evolution of flux yielded to a TeV period of $1610\pm58$ days (6\% probability){.

\begin{figure}[h!]
\centerline{\psfig{file=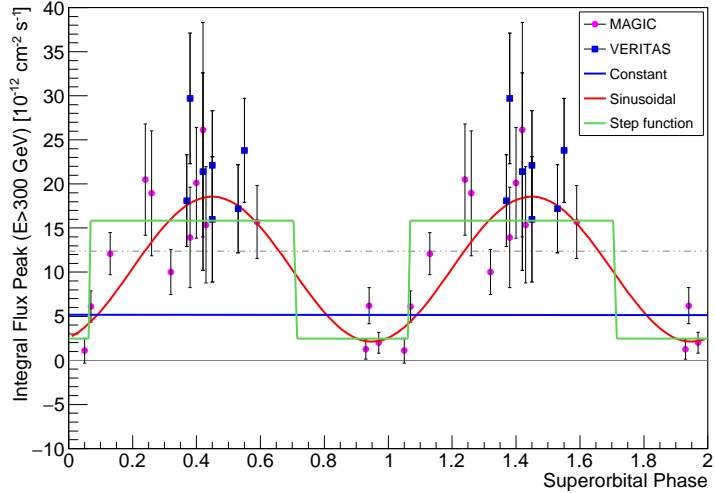,width=10cm}}
\caption{Peak of the VHE emission\cite{Ahnen2016} in terms of the super-orbital radio phase\cite{Gregory02}. Each point represents the peak flux emitted in one orbital period during apastron ($\phi$=0.5-0.75) measured by MAGIC (magenta dots) and VERITAS (blue squares). Fits with a sinusoidal (solid red line), step function (solid green line), and a constant (solid blue line) are also represented. As reference, we plotted $10\%$ of the Crab Nebula flux (dashed gray line) and zero level (gray solid line). \label{f1}}
\end{figure}


Furthermore, we have studied the possible correlation between the TeV flux measured by MAGIC and the H$\alpha$ parameters measured with the LIVERPOOL optical telescope, to search for (anti-)correlation between the mass-loss rate of the Be star and the TeV emission. No statistically significant correlation was found for the sample at orbital phases $\phi$ = 0.75-1.0 (phases of sporadic VHE emission\cite{lsifluxstates}). However, a stronger correlation might be blurred due to the fast variability of the optical parameters on short timescales compared to the long exposure times required by MAGIC.

\section{Search for VHE Emission from (high-mass) microquasars}

VHE gamma-ray emission in microquasars can be produced via leptonic\cite{Atoyan1999,BoschRamon2006} and hadronic \cite{Romero2003} processes. This emission could arise close from the binary, due to inverse compton (IC) of thermal photons or synchrotron photons. It could also be produced due to the interaction between the jets and the surrounding environment. 

\subsection{SS~433}

The eclipsing binary SS~433 is composed of a $\sim$10--20~M$_{\odot}$ BH \cite{Margon1984} orbiting ($P_{\rm orb}\sim13.1$~days) an A3-7 supergiant star. It displays supercritical accretion onto the compact object via Roche lobe overflow and it shows two baryonic relativistic jets which have a precessional period of $\sim$162.4 days. SS~433 is embedded in the radio shell SNR W50. The jets interact with the shell in the eastern and western blobs \cite{Goodall2011}. HE emission has been detected with Fermi-LAT\cite{Bordas2015}. In a leptonic scenario, VHE gamma rays could be produced via IC scattering of ambient photons and emission from the extended accretion disk \cite{Gies2002,Fuchs2006}. In addition, synchrotron-self Compton and relativistic Bremsstrahlung processes could also happen\cite{Aharonian1998}. In a hadronic framework, interactions of relativistic protons in the jets can produce gamma rays through $\pi^{0}$ decay \cite{Reynoso2008b}. However, since the binary is embedded in a thick extended envelope \cite{Zwitter1991}, the putative gamma-ray emission from the inner regions could be absorbed along $\sim$80\% of the precessional cycle \cite{Reynoso2008b}. The minimum absorption is expected to happen at phases {$\Psi_{\rm pre}$=0.9-0.1}. VHE emission can also be produced at the interaction regions between the jets and the nebula.


Multi-year observations (2006-2011) of SS~433/W50 were conducted in a joint campaign with the MAGIC and H.E.S.S. telescopes, during the phases of low absorption. Data analysis were performed following the standard procedure for each instrument (see \refcite{Aharonian2006b} for H.E.S.S. and {\refcite{Aleksic2016}} for MAGIC). No significant excess was detected neither from the central binary system nor from the interaction regions \cite{Ahnen2017}. A daily-basis analysis did not revealed flaring emission. Differential upper limits (ULs) were computed for each observatory and combined using a maximum-likelihood ratio test (see Fig.~\ref{f2}). This ULs are compatible with the model by Ref.~\refcite{Reynoso2008b} and the HE emission detected by Fermi-LAT \cite{Bordas2015}.

\begin{figure}[h!]
\centerline{\psfig{file=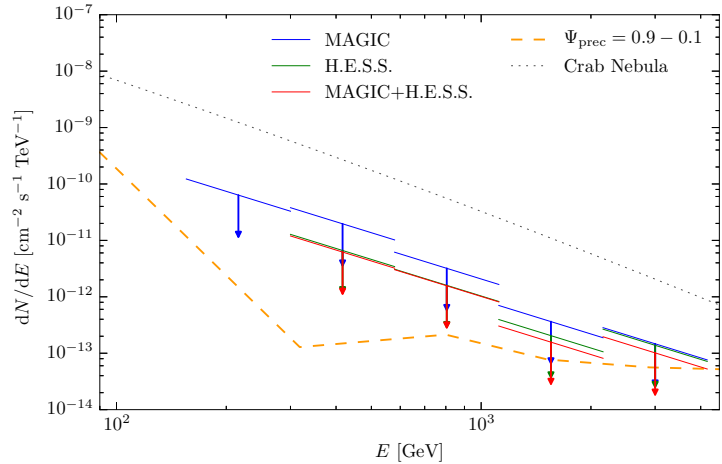,width=10cm}}
\caption{Differential flux ULs obtained by MAGIC (blue), H.E.S.S. (green) and a combination of both (red)cite{Ahnen2017} assuming a power-law with a spectral index $\Gamma = 2.7$. The predicted differential gamma-ray flux\cite{Reynoso2008b} for precessional phases $\Psi_{\rm pre} \in [0.9,0.1]$ is also displayed (dashed orange), together with the Crab Nebula flux, for reference. \label{f2}}
\end{figure}

In a hadronic framework\cite{Reynoso2008b}, with the ULs reported here, we can constrain the fraction of power carried by relativistic protons to be $q_{p}\lesssim\,2.5\times10^{-5}$. In a leptonic scenario, VHE emission could be produced in the interaction regions due to the presence of electrons with energies up to 50 TeV. However, they might lose most of their energy through synchrotron emission due to magnetic fields B $\gtrsim20-25\mu$G (according to the reported ULs), preventing an efficient gamma-ray production.

\subsection{Cygnus~X-1}

Cygnus~X-1 is the first identified stellar-mass black hole (BH) X-ray binary. The system is composed of a massive O9.7 Iab supergiant star \cite{Oroz2011} and a 25-35\,M$_{\odot}$ BH \cite{Ziolkowski2014}. It displays a highly-collimated one-sided radio jet which extends for 15 mas \cite{Stirling2001}. The system surrounded by a shell nebula composed of collisionally ionized gas which is detected in radio \cite{Gallo2005}. Cygnus~X-1 shows orbital periodicity in X-rays and radio and a 300-day super-orbital period detected in X-rays \cite{Rico2008,Zdziarski2011}. The binary displays the two principal spectral X-ray states of a BH transient: the hard state (HS), when the jet is steady and persistent (although some flaring might be present \cite{Fender2006}) and soft state (SS), when the jet is disrupted. 

Cygnus~X-1 has recently been detected at HE \cite{Zanin2016} with Fermi-LAT during HS. The emission is likely to be produced outside the corona, arising from the jets. Three transient episodes have also been detected with AGILE during HS and intermediate state \cite{Bulgarelli2008,Sabatini2010,Sabatini2013}. A hint of emission at 4.1$\sigma$ (post-trial) was observed with MAGIC in 2006 \cite{Albert2007} after 80 min of observations, simultaneously to a hard X-ray flare when the source was in HS and during superior conjunction.

MAGIC performed observations of the source for 100 hours between 2007 and 2014, mainly during HS. No significant excess was detected (for E$\ge$200 GeV) for steady, orbital or daily basis emission \cite{MAGICsubm}. The spectral energy distribution (SED) is shown in Fig.~\ref{f3}. An orbital-phase analysis did not show any orbital modulation during HS (83 hours of observation) neither during SS (14 hours). Considering the total power emitted by the jets during HS, we can set an UL on the conversion efficiency of jet power to VHE luminosity, which is constrained to be 0.006-0.06\%. With these results, we can rule out VHE emission from the jet large scale or from the interaction between the jet and surrounding medium above the sensitivity level of MAGIC, since it is not affected by gamma-ray absorption. However, particle acceleration up to TeV could still happen inside the jet in the binary region, but this emission would be below the MAGIC detection level. Hence, transient emission at binary scale (as the hint seen by MAGIC in 2006) can not be discarded. More sensitive instruments as CTA would be required to detect steady TeV emission.

\begin{figure}[h!]
\centerline{\psfig{file=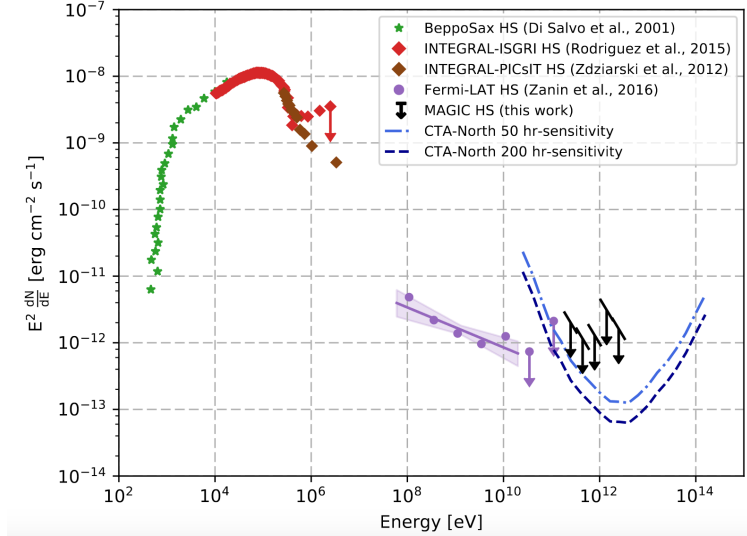,width=10cm}}
\caption{SED of Cygnus X-1 from X-rays \cite{diSalvo2001,Rodriguez2015} (green and red), HE \cite{Zanin2016} (violet) and VHE gamma-rays \cite{MAGICsubm} (black) during HS. The dashed blue lines correspond to the 50 and 200 hr sensitivity curves for CTA North. No statistical errors are drawn. \label{f3}}
\end{figure}

\section{Gamma-ray Burst alerts to follow up LMXBs: V404~Cygni}


The microquasar V404~Cygni is a low-mass X-ray binary (LMXB) composed of a 8-15\,M$_{\odot}$ BH and a $\sim$1\,M$_{\odot}$ companion star \cite{Casares1994,Khargharia 2010}. In June 2015, after 26 years in quiescence, the system displayed a major flaring episode in X-rays, reaching a flux about 40 times that of the Crab Nebula in the 20-40 keV energy band \cite{Rodriguez2015aa}. In the case of LMXBs, models predict TeV emission under efficient particle acceleration on the jets \cite{Atoyan1999} or strong hadronic jet component \cite{Vila2008}.
 
MAGIC observations were triggered by INTEGRAL alerts via Gamma-ray Coordinate Network (GCN), allowing an automatic and fast re-pointing of MAGIC to the burst position. The binary was observed for 8 non-consecutive nights during 10 hours, covering the strongest hard X-ray flares. The selection of the flaring times for the MAGIC analysis were performed running a Bayesian block analysis on the INTEGRAL light curve, assuming that the TeV flares were simultaneous to X-ray flares, which led to a data set of 7\,h. No significant emission was detected above 200 GeV and integral and differential ULs were computed \cite{MAGICinprep}. On the night of  26th June, the jet environment dramatically changed and a hint of detection (at 4$\sigma$) was seen by Fermi-LAT\cite{Loh2016}.  MAGIC had 1 hour of simultaneous observations with the Fermi-LAT excess, but no VHE signal was detected. The SED of V404 Cygni displaying the differential ULs for the complete set of observations and for that on 26th June are shown in Fig.~\ref{f4}.


\begin{figure}[ht!]
\centerline{\psfig{file=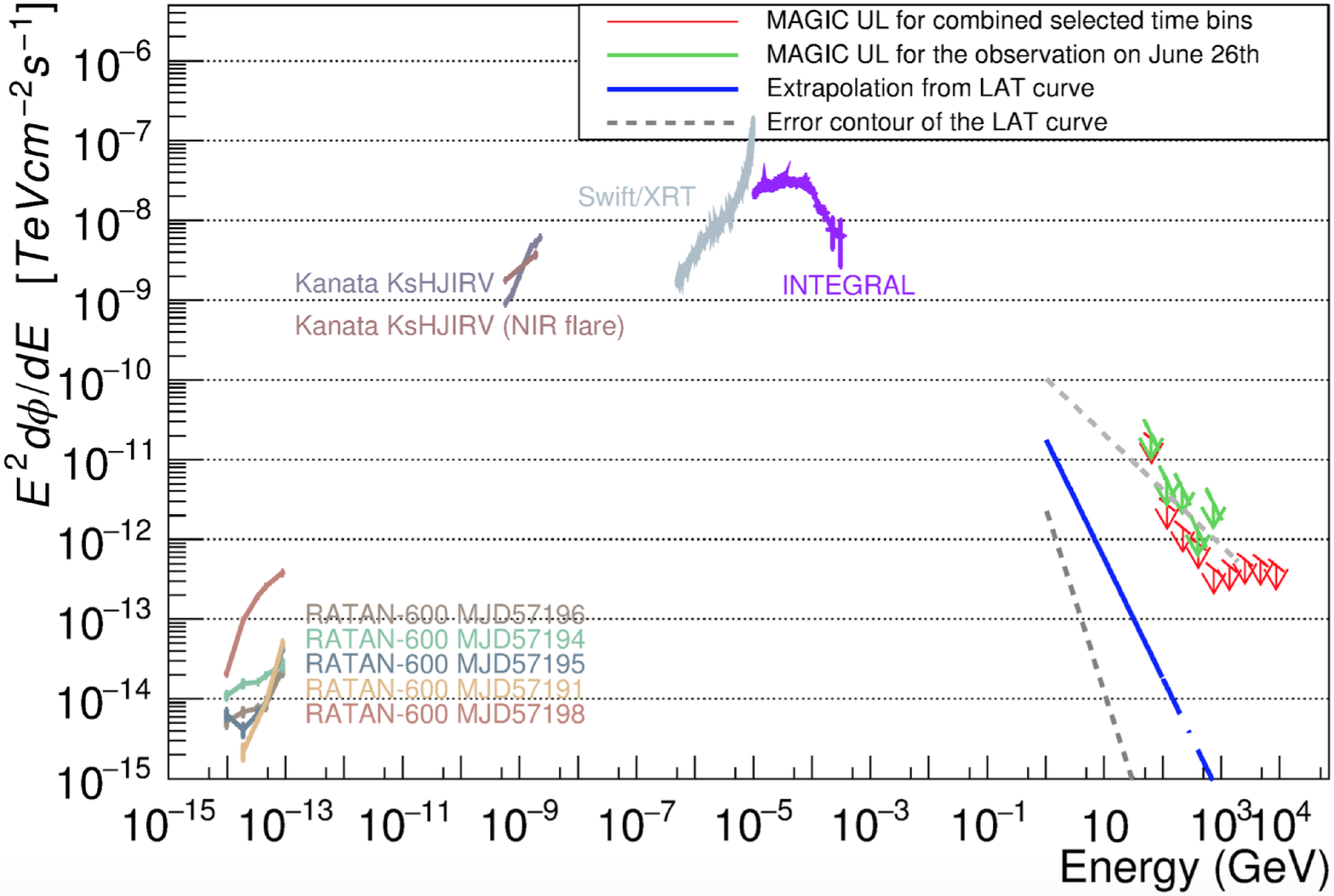,width=10cm}}
\caption{SED of V404 Cygni during the June 2015 major outburst. MAGIC ULs \cite{MAGICinprep} for the analysis of the combined Bayesian Block sample (red) and observations on June 26th (green) simultaneously taken with the Fermi-LAT hint \cite{Loh2016} are represented. The extrapolation of the Fermi-LAT spectrum is shown in blue with 1$\sigma$ contour (gray dashed lines). INTEGRAL X-ray data \cite{Rodriguez2015aa}, Swift-XRT\cite{Tanaka2016} and RATAN-600 radio data \cite{Trushkin2015} are also plotted. \label{f4}}
\end{figure}

If VHE gamma-rays are produced in V404 Cygni, they may annihilate via pair creation in the vicinity of the emitting region and strong gamma-ray absorption can occur at the base of the jets (r$\le1\times10^{10}$cm). At larger scales, VHE absorption is negligible, hence if VHE emission is produced it shall not be affected by pair production attenuation. In this case, we can constrain inefficient emission (0.003\%) in the jets. Our results suggest either a low particle acceleration rate inside the jets or not enough energetics.


\section{Summary}

MAGIC has performed an extensive campaign in the search for VHE emission from compact binary systems. We can summarize our latest results as follows:

\begin{itemlist}
 \item MAGIC discovered  super-orbital modulation in the VHE peak of LS~I~+61$^{\circ}$303 \cite{Ahnen2016} (compatible with other wavelengths) which may be explained due to changes in the mass-loss rate of the companion star.
 \item No significant excess was detected in SS 433, neither from the central binary nor from the interaction regions with the W50 nebula \cite{Ahnen2017}. We set constrains on particle acceleration efficiency and on the minimum value of the magnetic field.
 \item Cygnus X-1 was not detected at TeV and ULs  for steady, daily and separated X-ray states (including phase-folded analysis) were computed \cite{MAGICsubm}. Interaction between jet and medium is discarded, however transient emission is still possible.
 \item The LMXB V404 Cygni was observed during the June 2015 major outburst, but no emission was detected\cite{MAGICinprep}, which implies  inefficient particle acceleration or not enough energetics.
 
\end{itemlist}

\end{document}